\documentclass[preprint,showpacs,preprintnumbers,amsmath,amssymb]{revtex4}

\usepackage{graphicx}% Include figure files
\usepackage{dcolumn}% Align table columns on decimal point
\usepackage{bm}% bold math

\begin{document}

%\preprint{APS/123-QED}

\title{Intense Atomic and Molecular Beams via Neon Buffer Gas Cooling}

\author{David Patterson, Julia Rasmussen}
%\altaffiliation[Also at ]{Physics Department, Harvard University.}%Lines break automatically or can be forced with \\
\author{John M. Doyle}%
\email{doyle@physics.harvard.edu}
\affiliation{%
Physics Department, Harvard University
}%

\date{\today}% It is always \today, today,

\begin{abstract}

We realize a continuous guided beam of cold deuterated ammonia with
flux of $3 \times 10^{11}$ ND$_3$ molecules s$^{-1}$ and a
continuous free-space beam of cold potassium with flux of $1 \times
10^{16}$ K atoms s$^{-1}$. A novel feature of the buffer gas source
used to produce these beams is cold neon, which, due to intermediate
Knudsen number beam dynamics, produces a forward velocity and
low-energy tail that is comparable to much colder helium-based
sources. We expect this source to be trivially generalizable to a
very wide range of atomic and molecular species with significant
vapor pressure below 1000 K. This source has properties that make it
a good starting point for laser cooling of molecules or atoms, cold
collision studies, trapping, or nonlinear optics in
buffer-gas-cooled atomic or molecular gases.

\end{abstract}

\maketitle

\section{\label{sec:level1}Introduction}

Cold, slow, laser cooled atomic sources of atoms have been
laboratory standards for over a decade but only recently have cold,
slow sources of molecules been demonstrated\cite{bm03}. This is
largely because the complex level structure of molecules makes laser
cooling, the workhorse of atom cooling, very difficult for the vast
majority of molecules. (Laser cooling of molecules has yet to be
demonstrated, although a recent proposal indicates a path towards
this with some molecules\cite{junyemolecules2008}.) Demonstrated
techniques to produce cold, low velocity samples of molecules are
buffer-gas cooling\cite{weinstein1998}, slow molecule filtering from
an warm, effusive source \cite{elecfilter}\cite{nda03}, deceleration
of molecules from a seeded supersonic source using time-varying
electric fields\cite{bochinski2003}\cite{bm03}\cite{tarbutt:173002},
photoassociation of cold and ultracold atoms \cite{sage2005}, and
the formation of molecules from ultracold atoms using Feshbach
resonances \cite{kokkelmans2001}.  In general, the fluxes achieved
with these molecular sources have been far lower than those achieved
in cold atom sources.

\subsection{Buffer-gas Cooling and Direct Injection of Atoms and Molecules}

Cryogenic helium buffer gas cooling with has proven a versatile tool
for producing cold, dense stationary gases and high flux beams of
cold atoms and molecules\cite{weinstein1998}\cite{patterson2007}.
The species of interest is injected into a cell containing cold
helium gas where it thermalizes, producing a cold mixture with the
helium. Until recently, the primary methods of injection have been
laser ablation, \emph{in situ} electric discharge, and direct
introduction of hot molecular beams. Ablation and discharges have
proven to deposit far more energy into the buffer gas than is
necessary for cooling the injected species. For example, in laser
ablation of a typical metal, a laser pulse with an energy of
$\approx$ 10 mJ will result in $\approx 10^{12}$ buffer gas cooled
atoms. Assuming a heat of vaporization of 1000 K per atom, this
represents an efficiency of $10^{-4}$\cite{bmichniak}. Even in the
best cases, low efficiencies limit the available flux of cold atoms
or molecules, and generally requires experiments to be run in a
pulsed mode.  Directly injecting a hot molecular beam is very energy
efficient, but so far has been limited to peak input fluxes of $5
\times 10^{13}$ molecules s$^{-1}$ in 10 ms pulses.
\cite{campbell2007}.

Using helium buffer-gas cooling, we previously \cite{patterson2007}
produced a continuous beam of cold molecular oxygen, $^{16}$O$_2$
guided 30 cm from a cryogenic cell to a room temperature UHV
chamber. In that work, oxygen gas was flowed into the cold ($T
\approx$ 1.5 K) helium buffer gas through a $T \approx$ 60 K
capillary line. The oxygen cooled and then exited through a 4 mm
diameter aperture. The resultant beam was used to load a magnetic
guide at high molecular flux. In this work we present substantial
developments of this technique and demonstrate them by continuous
production of large fluxes of cold potassium atoms and deuterated
ammonia molecules. We employ a much hotter capillary ($T = 600$ K)
and use, in addition to helium, a neon buffer gas. We show that the
use of neon as a buffer gas leads to beams with more slow atoms or
molecules than helium-based beams, even though the neon gas is at a
significantly higher temperature (due to vapor pressure
constraints). This new variant of buffer-gas beam sources represents
a major departure from previous approaches and can be trivially
generalized to provide a continuous, high flux source of essentially
any species that has a significant vapor pressure at around 1000 K
or below.
%This source represents a possible starting point for
%loading magnetic or electrostatic traps, collisional studies, and
%nonlinear optics experiments.

%In this letter we present an extension of this cold buffer-gas beam
%source.  Using cold ($\approx 15 K$ Neon as the buffer gas, we have
%produced high flux (XXX Atoms/sec) cold (15 K) beams of alkali
%atoms, introduced into a cold cell via a capillary from a hot (600K)
%oven. We have in addition used a similar technique to produce a high
%flux (1e15 ND$_3$ s$^{-1}$ CHECK) unguided beam of state selected
%ND$_3$ at translational and rotational temperatures of $\approx$ 15
%K, and coupled this source into an electrostatic beam guide to
%produce with more modest flux of colder, state-selected molecules.

\section{\label{apparatus}Apparatus}
\begin{figure*}
\includegraphics{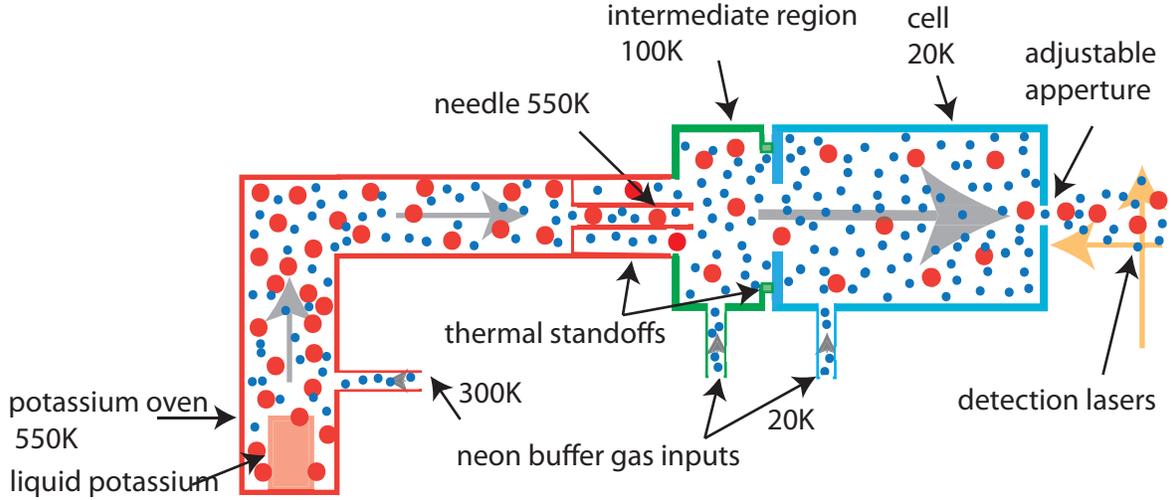}
\caption{Overview of apparatus. An oven supplies a hot (550K)
mixture of potassium and neon that flows through an intermediate
region (100K) to a cold tube ``cell''. (15K). The oven, intermediate
plate, and cell are separated by thermal standoffs to reduce heat
loads. A cold beam of potassium and neon exits an adjustable
aperture on the far side of the cell where its characteristics are
measured by absorption spectroscopy. \label{apparatusfig}}
\end{figure*}

  Our potassium beam apparatus is centered around a cold tube (the ``cell''), $T$ = 6-15K, into which
we flow target atoms or molecules and inert gas, as depicted in
Figure \ref{apparatusfig}.
  This cell is the coldest part of the apparatus and is 1 cm in diameter and 8 cm long.
A hot (up to 600 K) gas mixture of neon and potassium flows into one
end of the tube. The mixture cools to the temperature of the tube
walls as it flows down the tube and out into vacuum, realizing the
beam.

In more detail, a hot (up to 600 K) copper tube is attached via a 2
cm long stainless steel tube to a copper plate (about 100 K), which
in turn is anchored to the first stage (40 K) of a closed cycle
pulse-tube refrigerator\cite{pulsetube}. This plate absorbs the bulk
of the radiative and conductive heat load from the hot input tube,
as well as possibly precooling the gas mixture The tube continues
via a 1 cm long, 1 cm diameter, thermally isolating section of
epoxy/fiberglass composite and enters the cold cell through a 6 mm
aperture in a copper plate. The cold cell is thermally connected to
the second stage (6 K) of the pulse tube refrigerator and has an
adjustable aperture (typically 3 mm diameter) at the far end. Total
heat loads on the cryorefrigerator from the apparatus are about 40 W
at the first stage and 1 W at the second stage, which coincides both
with the cooling power of the pulse tube, and with typical heat
loads in a nitrogen/helium liquid cryogen cryostat. The total
distance between the end of the hot copper tube and the entrance to
the cell is 1.5 cm. Our current maximum running temperature of 600 K
is limited by this 40 watt heat load on the first stage of the
cryocooler, and could be significantly increased by better thermal
isolation, or the construction of an additional intermediate cooling
plate at 300 K.

Buffer gas can be introduced -- in combination or separately -- at
the oven (600 K), the first copper plate (100 K), or the cell (15
K).  The temperature of the gas mixture at the exit of the cell is
measured to be $\approx$ 15 K regardless of where the gas is
introduced (see section \ref{thermalizesection} below); the flux of
cold potassium is maximized when the buffer gas flow through the
oven dominates over the flow into the plate or cell, likely because
this flow transports the comparatively low pressure potassium (1
torr at 600 K) down the long hot copper tube and cooling occurs
directly within the cell.  Our potassium flux is limited by the
vapor pressure of potassium at our current maximum running
temperature of 600 K, suggesting that fluxes could be substantially
higher if we were using a species with a higher vapor pressure. The
oven and transport tube could be run at higher temperature with
straightforward engineering improvements to the thermal isolation.

\section{Thermalization and Loss to the Cell Walls}
\label{thermalizesection}

The key physical process that leads to production of cold potassium
is the flow and thermalization of the initially hot potassium/buffer
gas mixture within the cell. In order to realize a cold, high flux
beam, the experimental conditions generally need to satisfy two
competing conditions. The gas mixture must thermalize with the cold
walls, and the mixture must exit the cell (a tube) before much of
the potassium diffuses to the walls, where it has a high probability
of permanently sticking.

  It is not clear \emph{a priori} that it is possible to achieve
thermalization without significant atom loss.  Assuming a tube of
radius $r$, a neon density of $n_{Ne}$, a gas temperature $T$, and
hard sphere elastic cross sections of $\sigma_{Ne-Ne}$ and
$\sigma_{K-Ne}$ respectively, the characteristic time $\tau$ for the
gas to thermalize with the walls is

\[\tau = r^2 n_{Ne} \sigma_{Ne-Ne} \left(m_{Ne} / 2 k_B T\right)^{1/2}
\]

Assuming that $n_{Ne} \gg n_{K}$, the diffusive loss of potassium to
the walls will occur on a time scale $\tau_{loss}$:

\[\tau_{loss} = r^2 n_{Ne} \sigma_{K-Ne} \left(m_{Ne} / 2 k_B T\right)^{1/2}
\]

%Note that the mass of the potassium does not enter this expression.

These expressions imply that in order to thermalize with only order
unity loss in atom flux, we need $\sigma_{Ne-Ne} \lesssim
\sigma_{K-Ne}$. Although we did not make accurate density
measurements in our cell under actual running conditions due to the
high ($>$ 100) optical density, we have evidence that loss to the
cell walls is minimal; we have been unable to find a regime where
the temperature of the beam differs significantly from the measured
temperature of the cell walls.  We therefore speculate that
$\sigma_{Ne-Ne} < \sigma_{K-Ne}$, but note that this conclusion and
the sensitivity of the overall cooling process on ratio of the
elastic cross section between buffer atoms to that between buffer
atoms and target species
 remains open to both theoretical and experimental investigation.

%\subsubsection{scratchwork}
%\[
%  mean velocity = sqrt{2k_B T / m}
%  mean free path l = 1 / n \sigma
%  mean free time = l / v = sqrt(m/ 2 k_B T n^2 \sigma^2)
%  number of steps = (r/l)^2
%  total time = r^2 / l v = r^2 n \sigma sqrt(2 m / 2 k_B T)
%\]

\section{Neon As a Buffer Gas}

In order to achieve the lowest possible temperature, the
overwhelming majority of buffer gas cooling experiments to date have
used helium as the buffer gas. With a sufficiently high vapor
pressure at temperatures as low as 300 mK (for $^3$He), compared to
a lower limit of $\approx$ 12 K for neon, helium is the natural
choice when low temperature is the only figure of merit.

Both magnetic and electrostatic traps have depths of at most a few
K. It is therefore natural to maximize the flux of molecules with a
total energy less than a given energy $E_{trappable} \approx 1$ K,
assumed
  here to be less than $T_{cell} \approx 4-20 K$.  Somewhat counterintuitively, by this metric a hydrodynamic \cite{patterson2007}
beam using 15 K neon as the buffer gas can substantially outperform
a similar beam using 4 K helium.

\begin{figure*}
\includegraphics{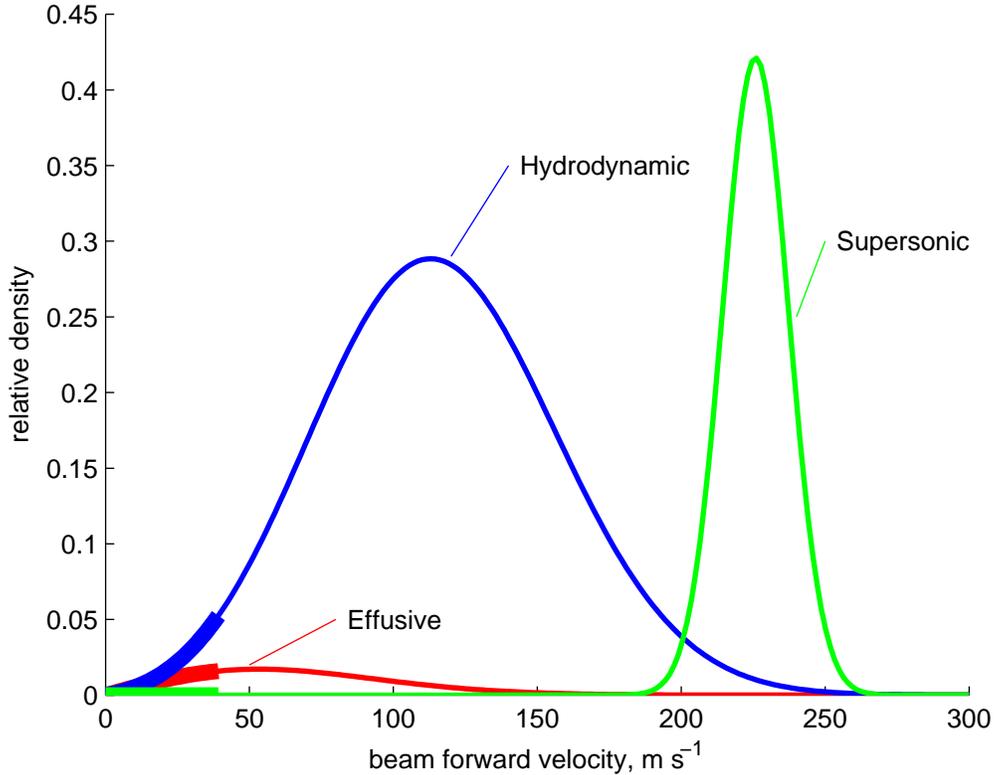}
\caption{Simulated forward velocity distributions for buffer gas
beams in various regimes. As a mixture of potassium and neon buffer
gas emerges from the nozzle, a typical potassium atom collides with
neon atoms as it moves away from the nozzle aperture. Qualitatively,
effusive beams suffer zero collisions, ``hydrodynamic'' beams a few
collisions, and supersonic beams many collisions, remaining fluid
dynamic for many aperture diameters. The beams demonstrated in this
letter are hydrodynamic, but generally not supersonic [blue
curve].\label{regimesfig}}
\end{figure*}

 Consider a beam consisting of a small fraction of potassium ``impurities'' of mass $m_K$
mixed with a much larger flow of a nobel gas of mass $m$. The
mixture is assumed to thermalize to a temperature $T$ before it
exits the nozzle.  Depending on the number of collisions within and
just outside the nozzle, such a beam can operate in one of three
regimes:
\begin{itemize}
  \item \emph{effusive}, characterized by $N < 1$ collision.  In such
  beams, only the atoms in the cell which happen to diffuse through
  the small nozzle aperture are emitted into the beam.  These beams
  have a forward velocity of order $\bar{v} \approx (2k_BT/m_K)^{1/2}$ and relatively low flux.

  \item \emph{hydrodynamic}, characterized by $1 < N \lesssim 100$ collisions;
 the flow through the nozzle is hydrodynamic, but
  isothermal.  These beams typically have a large flux enhancement compared to effusive beams due to hydrodynamics
within the cell, as described in \cite{patterson2007}. Hydrodynamic
beams have a modestly boosted mean forward
  velocity of $\bar{v} \approx (2k_BT/m)^{1/2}$.

  \item \emph{fully supersonic}, characterized by $N > 1000$ collisions
  in and outside the nozzle.  In this regime, there is significant
  adiabatic cooling as the gas expands into the vacuum; the forward velocity distribution
 is sharply peaked around a velocity $\bar{v} \gtrsim 2 \times (2k_BT/m)^{1/2}$.

\end{itemize}
Figure \ref{regimesfig} shows the forward energy spectrum to be
expected in each of these regimes.

\begin{figure*}
\includegraphics{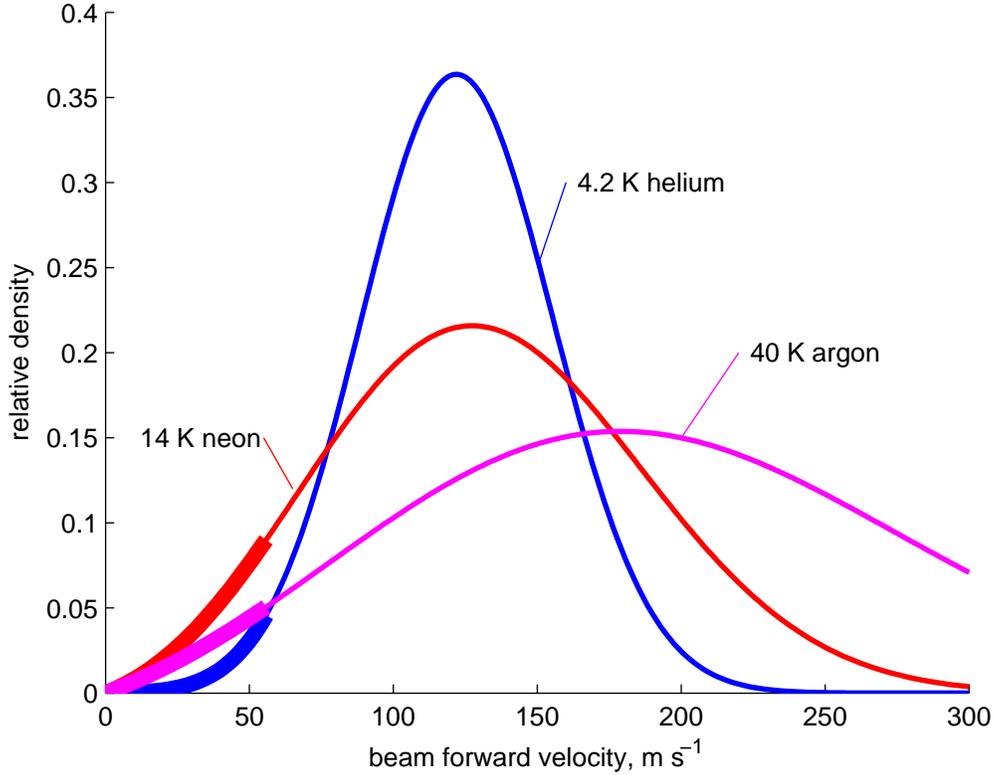}
\caption{Theoretical energy spectrum for hydrodynamic beams using
4.2 K Helium, 14 K Neon, or 40 K Argon as a buffer gas.  The portion
of the distribution with an energy of less than 4K is shown in bold.
Neon is clearly an attractive choice for experiments interested in
maximizing the flux of very cold atoms or molecules.
\label{variousgastheoryfig}}

\end{figure*}

  In order to take advantage of large flux enhancements,
  we run our beams in the hydrodynamic regime described above (blue
curve in figure \ref{regimesfig}). Figure \ref{variousgastheoryfig}
shows the simulated forward velocity distribution for such a source
using 4.2 K helium, 15 K neon, and 40 K argon as the buffer gas.  It
is clear that despite the higher temperature required for a neon or
argon beam, there are more cold, low energy atoms in the neon beam
than in the helium beam.

  Using neon instead of helium as the buffer gas carries several
important technical advantages as well.  Most notably, neon has a
negligible vapor pressure at our cryostat's base temperature of 5 K,
meaning that any cold surface becomes a high speed cryopump.  The
vacuum outside the cell can be maintained at a high level ($1 \times
10^{-6}$ torr in the main chamber, $< 10^{-8}$ torr in a
differentially cryopumped ``experiment chamber'') even with a high
flow of order $10^{19}$ neon atom s$^{-1}$ through the cell.
  In future experiments, simple
shutters could be used to rapidly close off this differentially
pumped chamber, which we believe will rapidly achieve an extremely
high vacuum. The higher temperature means the cell can tolerate
substantially higher heat loads. The principle disadvantage of neon
is that the higher temperature leads to less rotational cooling in
molecules than that available from 4 K helium.
%These advantages can be compared
%to neon's principal disadvantage: that we are forced to hold the
%temperature of the cell above 12 K in order to maintain a
%significant buffer gas density.

\section{\label{kresults}Potassium Results}
\begin{figure*}
\includegraphics{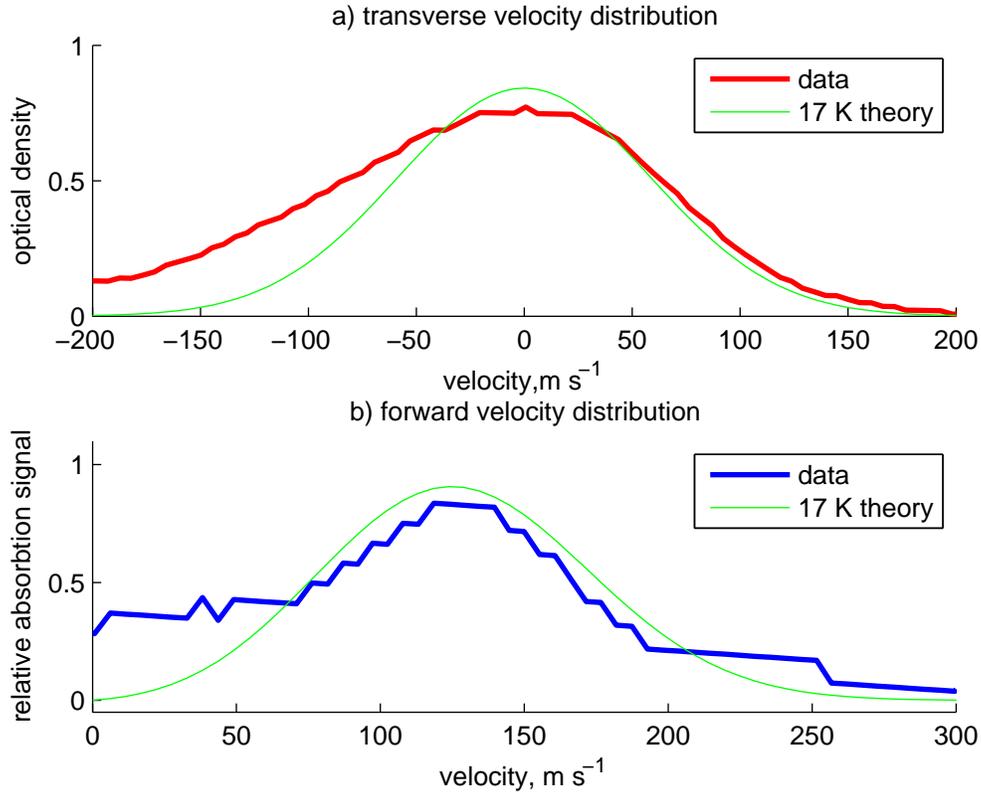}
\caption{Measured transverse [a] and forward [b] velocity
distribution of a high flux beam of potassium.  The green curves
represent the expected distribution from a hydrodynamic beam with
$T_{gas} = 17 K$.  The amplitude is the only free parameter in these
fits.  The left side of each plot is somewhat distorted by the
nearby $F=0$ hyperfine line.  This data represents a beam with a
potassium flux of $1 \times 10^{15}$ atoms / second; we have
observed fluxes as high as $1 \times 10^{16}$ atoms / second, but
they are not shown here because the high optical density ($\approx
8$) makes their interpretation less straightforward.}

  \label{potdatafig}
\end{figure*}

  A cold, high flux beam of potassium was characterized to study this new beam system.
  Two lasers are used to characterize density and velocity distribution in the atomic beam via absorbtion spectroscopy.
  The first, used to measure
  the beam flux and transverse velocity distribution, passes through
  the center of the combined neon/potassium beam 2 cm
  away from the nozzle.
% which we believe to be the approximate location of the
 % ``zone of freezing'' - the typical location of the final collision with a neon
 % that a potassium atom undergoes.
The second laser propagates
  parallel to and slightly below the atomic beam, intersecting
  the edge of the beam.  This laser is used to measure the forward
  velocity distribution of the beam.
  Figure \ref{potdatafig} shows typical measured velocity distributions.

  Under typical running conditions (potassium oven and injection needle
  heated to 580 K, the cell held at 17 K, 110 sccm of neon
  buffer gas flowing), we measure an optical density of about 5 in the transverse laser path,
  and a mean forward velocity of 130 m/sec, in excellent agreement
  with a simple simulation of a hydrodynamic beam at 17 K.
  These measurements imply a density of about $8 \times 10^{11}$ potassium cm$^{-3}$ with an atomic beam diameter of about 1 cm
   and a total beam flux of $1 \times 10^{16}$ atom s$^{-1}$. The beam divergence is measured to be
about 0.7 steradians, consistent with our model of beams in this
hydrodynamic regime.

   %Assuming that \emph{all} the atoms in
   %the cell tube are swept into the beam, this measurement implies
   %a lower bound on the density in the cell of about $10^{13}$, with
   %a lower bound on the optical density of about 50.

%\subsubsection{scratchwork}
 % Absorbtion cross section (natural) = $\lambda^{2} / 2 \pi$.
 % Natural linewidth is 7 MHz.  I have 200 MHz. = 30 x as much. So
  %effective cross section is $1 \times 10^{-9} / 30 = 3 \times
 % 10^{-11}$.

  %$OD = nL\sigma$, so $n = OD / L \sigma$
  %so $n = 4 \times 10^{11}$ per state or $8 \times 10^{11}$
  %Assume a cone of about 30 degrees (valid by absorbtion spec),
  %which gave $L \approx 1$ cm, and $v = 13000$ cm/sec, then we have
  %about $10^{16}$ /sec

\begin{figure*}
\includegraphics{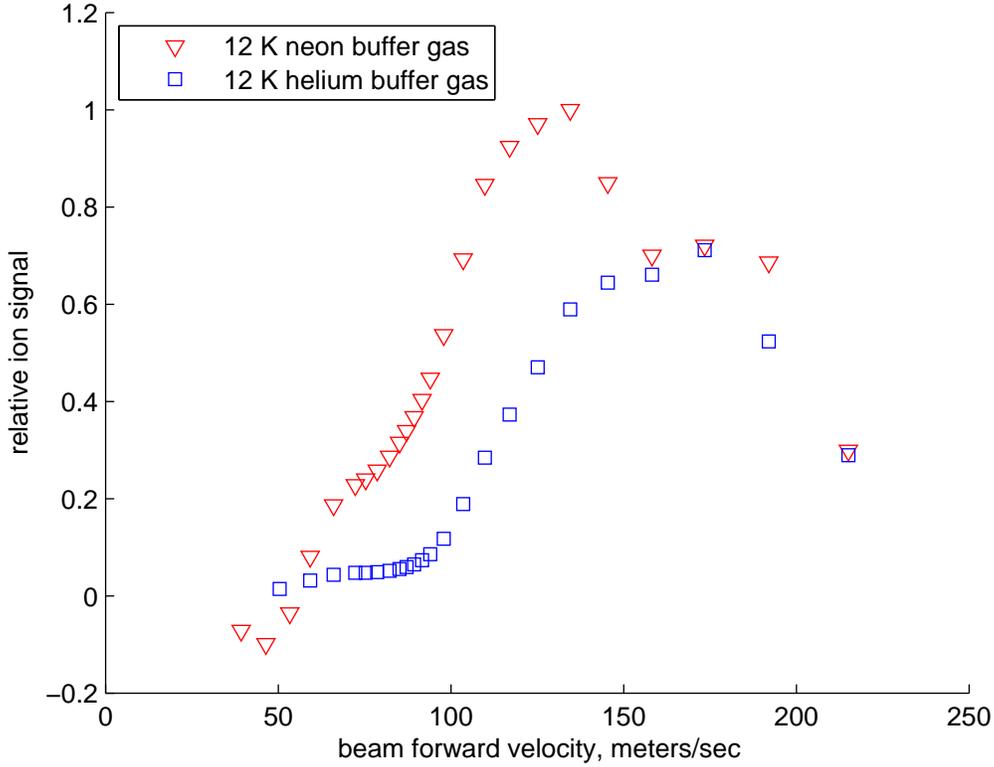}
\caption{Measured forward velocity distribution of a guided ND$_3$
beam using 12 K helium [blue square] and 12 K neon [red triangle] as
a buffer gas. The beam using neon contains substantially more
low-velocity molecules. Molecules with a velocity higher than 200 m
s$^{-1}$ are moving too fast to be guided by this bent electrostatic
guide (bend radius $r$ =70 cm.)  The red data [neon] represents a
guided beam of about $2 \times 10^{10}$ ND$_3$ molecules s$^{-1}$ at
a mean energy of $\approx$ 20 K.}

  \label{heliumneondatafig}
\end{figure*}

\begin{figure*}
\includegraphics{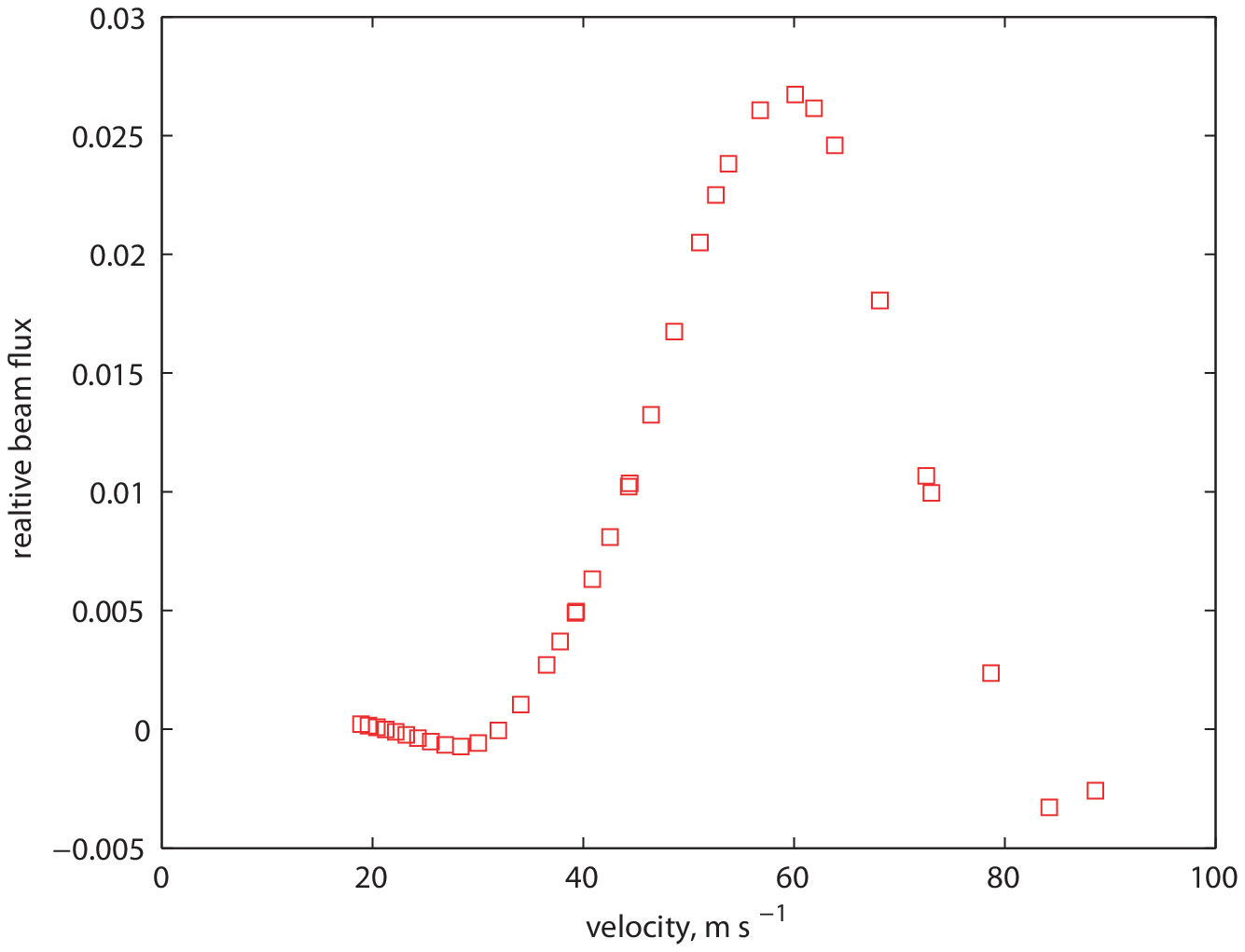}
\caption{Velocity distribution at the output of a 70 cm long
electrostatic guide with a bend radius of 8 cm.  The output of this
guide is completely separated from the neon buffer gas, and
represents a pure, state-selected, low energy beam.  This data
represents $3 \times 10^{8}$ ND$_3$ molecules s$^{-1}$, with a mean
energy of about 4 K.\label{ND3 data}}

  \label{trappic}
\end{figure*}

\section{Applications}
\subsection{Demonstrated: Loading an electrostatic beamguide}
\label{ammoniaresults}

 It is natural for a variety of experiments to want a cold beam of molecules
that has been separated from the helium or neon buffer gas.  As an
example application of this source, we have demonstrated loading a
cold beam of ND$_3$ into a curved electostatic guide; the output of
such a guide consists of pure, state-selected ND$_3$ molecules. (We
note that after our ND$_3$ experimental work was completed we
learned of similar results presented in \cite{buurenrempe2008}).  In
addition, the guide acts as a filter that passes only molecules
moving slowly enough to be guided around the bend.  Such a guide can
be used to bring the beam into a differentially pumped region,
either cryogenic [this work] or at room
temperature\cite{buurenrempe2008}\cite{patterson2007}

The apparatus for our guided ND$_3$ work is simpler than for our
high flux potassium beam. Specifically, the ND$_3$ is fed directly
into the cell and the capillary temperature is a lower $T =$ 280 K.
Cold ND$_3$ molecules exit the cell and after a short gap of 30 mm,
enter an electrostatic guide.  ND$_3$ molecules exiting the guide
were detected via 2+1
  REMPI using a 6 mJ pulsed laser at 317 nm.
The guide consists of six 1 mm diameter stainless steel rods with a
center-center spacing of 2 mm.  The guide entrance is positioned 30
mm from the exit aperture of the cell, slightly beyond the expected
position of the last collision between a typical ND$_3$ molecule and
buffer gas atom.  The position of this last collision is velocity
dependent, with slow molecules suffering more collisions.  We
therefore speculate that the dominant loss mechanism for the very
slowest molecules in the guided velocity distribution is collisions
with buffer gas atoms in the first few cm of the guide.

The guide is constructed from alternating rods held at 5 kV and
ground, producing a linear hexapole field. The guide rods were held
in place with Ultem brackets; the entire assembly was mounted to the
6K cold plate.  No detectable
 ($<$ 0.1 $\mu$A) leakage current flowed along the
surface of the Ultem brackets despite the fact that few precautions
were taken to avoid such currents.  We speculate that such surface
currents are suppressed at low temperatures.  The guide is 70 cm
long and describes a complete loop with a bend radius of 8 cm. The
first 30 cm of the guide can be switched on or off independently
from the
  rest of the guide.

In a typical experimental run, the first 30 cm of the guide is
turned off and the remainder of the guide is charged.  The first
section is then turned on, and, after a variable delay, the pulsed
REMPI laser is fired, measuring the density of molecules a few cm
away from the guide exit.  The velocity distribution in the guide,
shown in figure \ref{ND3 data}, is inferred by varying this delay.
We can only present a conservative lower bound on the guided
molecule flux as absolute density measurements are notoriously
unreliable in MPI experiments; our quoted flux is based on very
conservative comparisons with similar, calibrated experiments on
ND$_3$ done in other groups\cite{bethlem:491}. We find a lower bound
of $3 \times 10^{8}$ guided molecules per second with an average
energy corresponding to a temperature of about 4 K.

The ND$_3$ beam was initially demonstrated using both helium and
neon as a buffer gas. Figure \ref{heliumneondatafig} shows the
measured velocity distribution of a guided beam of ND$_3$ using
helium and neon as buffer gas under otherwise similar conditions.
The guide used in this work has a significantly larger (70 cm)
radius of curvature, leading to a much higher velocity cutoff in the
beam guide output velocity distribution. Significantly more slow
molecules are produced in the beam using neon buffer gas.

\subsection{Next step I: Collisional studies}
  There is substantial interest in studying both elastic and
  inelastic collisions of molecules at low energies\cite{bm03}.
  The guided
  output of the deuterated ammonia beam described above represents a pure, rotationally cold,
  state-selected, slow molecular beam. In addition, a particular velocity class can be selected by controlled switching of the guide.  Such a
  beam could be crossed with a similar beam, a
  laser cooled atomic beam, a trapped sample, or a cold thermal beam of hydrogen,
  helium, or neon, providing a ``lab bench'' to study
  cold collisions with energies of a few cm$^{-1}$.
  Such a system would provide an attractive set of ``knobs'' to the
  experimenter - collision energy, incoming rotational state, and
  external fields can all be controlled. Even a small fraction of
  the beam being scattered inelastically into high field seeking states would be detectable,
since the output of the beam guide consists of very pure low field
seeking states.

\subsection{Next step II: General loading of traps}
  Low velocities and high fluxes make this source an attractive first step for loading
electrostatic or magnetic traps.  In order to load atoms or
molecules into a conservative potential, energy must be removed from
the particles while they are within the trap volume.  Demonstrated
methods to load traps from beams include using light
forces\cite{chu1998}\cite{helmersonlasercooling}, switching
electric\cite{bethlem:491} or magnetic\cite{Raizenslowing} fields,
and buffer gas loading \cite{campbell2007}. Chandler et.
al.\cite{evc03} have proposed loading traps via collisions with
another beam, and it should also be possible to optically pump slow
moving atoms or molecules directly into trapped states\cite{ssh01}.
Recent proposals\cite{junyemolecules2008} suggest molecules as well
as atoms could be laser cooled from a source like ours, and all
other methods listed here are trivially generalizable to molecules.
All of these methods would benefit from starting with our bright,
cold, slow moving source.

\subsection{Next step III: Nonlinear optics in the buffer gas cell}
  There has been recent interest\cite{Budker2008}\cite{TaoHong2008} in studying
  non-linear opticals effects in cold buffer gas cells.  There is theoretical\cite{Budker2008} and experimental\cite{hatakeyamacoldrb}
  evidence that decoherence times between ground-state magnetic sublevels in
alkali atoms are very long.  To date, only cells loaded via laser
ablation, and in one case light induced
desorbtion\cite{hatakeyamalightdesorbtion},
  have been used for this work. Although substantial optical
  densities have been achieved, such experiments tend to be plagued
  by anomalous losses at high buffer gas pressures, limiting
  lifetimes to $<$ 100 ms.  Ablation is a violent and
  difficult to control process which is known to produce both
  clusters\cite{croab} and momentary dramatic heating\cite{bmichniak} in buffer
  gas cells. We believe that continuously loading cold vapor
  cells directly from an oven will not only provide
  higher density samples, but also do so controllably in a clean
  environment.  Possible demonstrations include improved EIT\cite{TaoHong2008},
  guiding light\cite{MaraConfinedLight2008}, and high resolution magnetometry\cite{BudkerNonlinearReview}.
We have already achieved in the experiments described above
continuous in-cell potassium optical densities much greater than
100.
\section{Conclusion}
  We have demonstrated an extremely high flux and moderately cold
  atomic beam of potassium using a neon buffer gas. We believe this source can be trivially generalized to any
  atomic and many
molecular species that can be produced in a gas at
  temperatures up to 600 K and perhaps as high as 1000 K.  The demonstrated cold flux of potassium represents a new benchmark for cold atom fluxes.
In addition, a ND$_3$ from such a source has been coupled
  into an electrostatic beam guide. A broad range of possible
  applications with such guided molecules includes collision studies, trap loading and nonlinear optics studies.

This work is supported by the National Science Foundation (Grant
0551153).

\bibliography{doylenjop}% Produces the bibliography via BibTeX.

\end{document}